\documentclass[aps, amssymb, amsmath, superscriptaddress, prl, twocolumn, showemail]{revtex4-2}
\usepackage{graphicx}
\usepackage{color}
\usepackage{amsmath}
\usepackage{enumitem}
\usepackage{amssymb}
\usepackage{hyperref}
\usepackage{cancel}
\usepackage{ulem}
\usepackage{multirow}

\usepackage{pifont}

\newcommand{\be}{\begin{equation}}
	\newcommand{\ee}{\end{equation}}
\newcommand{\bea}{\begin{eqnarray}}
	\newcommand{\eea}{\end{eqnarray}}
\newcommand{\p}{\partial}

\renewcommand{\vec}[1]{{\mathbf #1}}
\newcommand{\veps}{\varepsilon}
\newcommand{\etal}{\eta_H}

\renewcommand\vec[1]{\ensuremath\mathbf{#1}} 

\usepackage{amsfonts, relsize, color}
\usepackage{graphicx}
\usepackage{color}
\usepackage{comment}

\usepackage{xcolor}
\hypersetup{
	colorlinks,
	linkcolor={red!50!black},
	citecolor={green!50!black},
	urlcolor={blue!50!black}
}
\usepackage{booktabs} 
\begin{document}
	

\title{Viscochiral Transport: Chiral Selection of Hydrodynamic Vortices by Berry Curvature}

\author{Archisman Panigrahi}
\email{archi137@mit.edu}
\thanks{The two authors contributed equally}
\affiliation{Department of Physics, Massachusetts Institute of Technology, 77 Massachusetts Avenue, Cambridge, MA 02139, USA.}

\author{Khachatur Nazaryan}
\email{khachnaz@mit.edu}
\thanks{The two authors contributed equally}

\affiliation{Department of Physics, Massachusetts Institute of Technology, 77 Massachusetts Avenue, Cambridge, MA 02139, USA.}

\begin{abstract}


We predict a new \textit{viscochiral regime} of electronic transport in which spatially varying Hall viscosity selects vortical flow patterns. Although uniform Hall viscosity cannot alter incompressible bulk flow, its spatial gradient redistributes vorticity, amplifying vortices in one chamber while suppressing the vortex in the other. We further determine the underlying mechanism to be generic to recirculating flows and insensitive to the details of the device geometry. The paper maps the resulting phase diagram and shows that the regime is experimentally accessible in valley-polarized bilayer graphene.

\end{abstract}
\date{\today}

\maketitle

\textit{Introduction---}
In recent years, there is a surging interest in the hydrodynamic regime of electronic transport, where momentum-conserving electron-electron collisions dominate over momentum relaxation to scattering by phonons and impurities, enabling collective flow at mesoscopic scales~\cite{Gurzhi1968,NarozhnyGornyi2017,LucasFong2018,FritzScaffidi2024, NazaryanLevitov2024, LevitovFalkovich2016}. 
As there is no momentum relaxation of flow in the bulk, the flow in the hydrodynamic regime features Poiseuille-like current profiles, negative nonlocal resistance, and vortical flow patterns~\cite{LevitovFalkovich2016,Bandurin2016,Bandurin2018,AharonSteinberg2022}. In contrast, in Ohmic regime the current flow is governed by a potential and lacks vortices~\cite{NazaryanLevitov2024,Zhang2026, Egorov_transport_2026}.
Recent scanning-magnetometry experiments have visualized both channel flow and chamber vortices in a dual-gated bilayer-graphene device whose geometry is simultaneously sensitive to laminar and recirculating flow~\cite{Zhang2026}. These developments make electron hydrodynamics a natural arena to explore how broken time-reversal symmetry reshapes viscous flow.

A particularly important non-dissipative response of time-reversal-broken fluids is odd (Hall) viscosity, first identified in quantum Hall fluids~\cite{Avron_viscosity_1995} and later formulated for two-dimensional hydrodynamics in electronic systems~\cite{Avron_odd_1998,Fruchart2023,RaoBradlyn2020}. In electron systems this response can be generated by Berry curvature and internal orbital angular momentum of Bloch wave packets~\cite{Hasdeo2021,SrivastavaMukerjee2025}, following the observation that the Hall viscosity is a geometric response tied to the intrinsic angular momentum of the state~\cite{Read_non-abelian_2009}. An experimentally relevant question is 
how it can control flow patterns in realistic device geometries. It was shown that the odd viscosity affects the nature of Coulomb drag between layers~\cite{Zverevich_hydrodynamic_2025}. In the context of hydrodynamic flow in a single device, a uniform odd-viscosity coefficient does not alter the velocity field of an incompressible flow at all~\cite{Ganeshan_odd_2017,RaoBradlyn2020,RaoBradlyn2023,Kirkinis_null-divergence_2023}. In a charged fluid it can still be accessed via the electrochemical potential, and existing proposals include measuring the potential profile near current-injecting contacts, in nonlocal resistance measurements and measurements of the Hall angle in Corbino geometries~\cite{Scaffidi_hydrodynamic_2017, Delacretaz_transport_2017, Holder_unified_2019}, experimentally demonstrated in Ref.~\cite{Berdyugin_measuring_2019}. By contrast, once the odd-viscosity coefficient varies in space, its gradient produces a directional body-force term that can bias circulation and break geometric symmetries of the flow. Since this work models a charged electronic fluid, we consider the incompressible limit throughout. Odd-viscous effects are known to be more pronounced in compressible flows~\cite{Avron_odd_1998}, where the effects on transport survive even for spatially uniform odd viscosity.

\begin{figure}
    \includegraphics[width=0.9\linewidth]{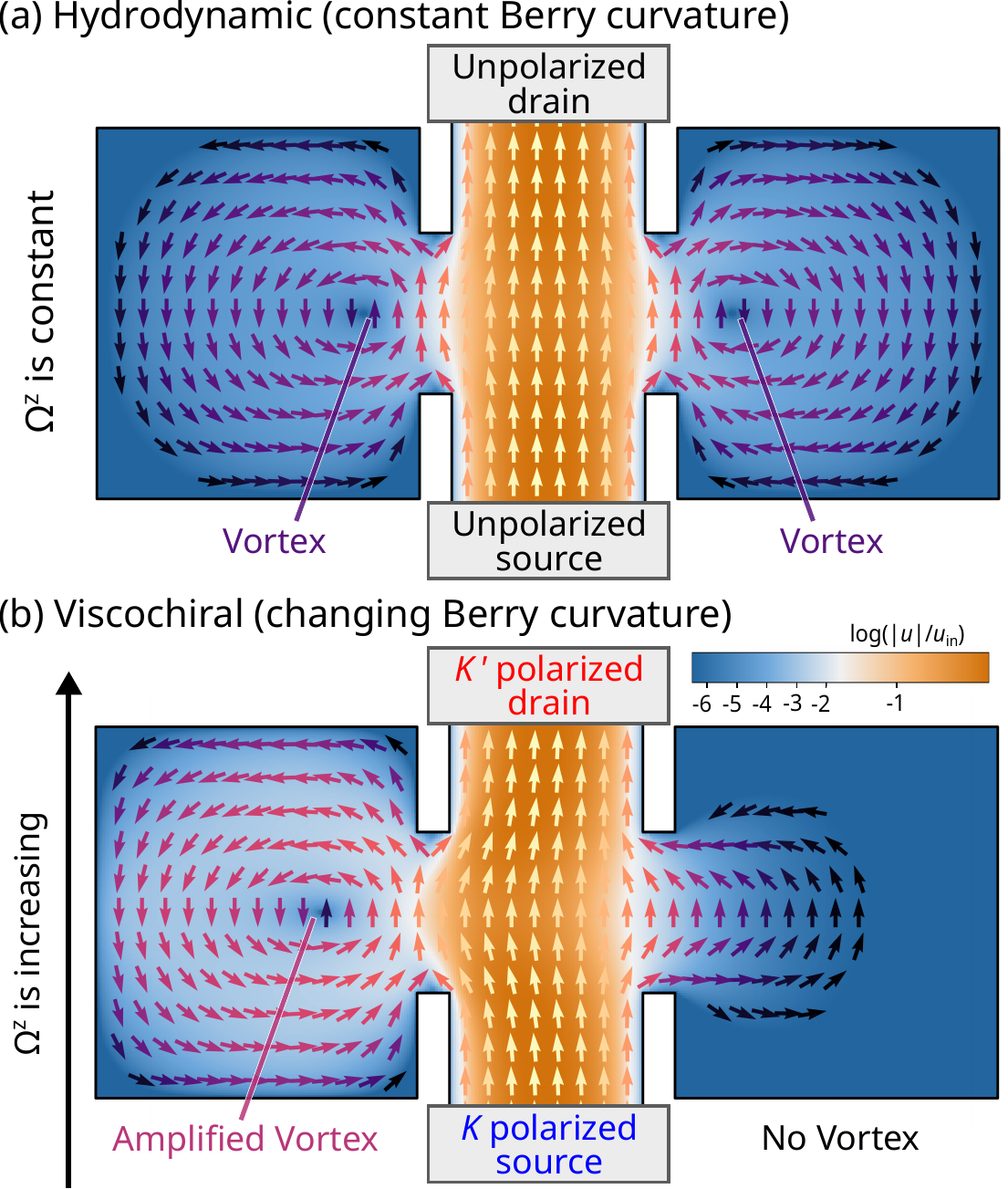}
    \caption{ 
    Chiral vortex selection by a gradient of Berry-curvature ($\Omega^z$) (and thus of Hall viscosity) in a two-chamber device.
    (a) For constant Berry curvature, the two chambers are equivalent and the
    hydrodynamic flow forms a pair of vortices with mirror symmetry.
    (b) When the Berry curvature varies across the device, the induced
    odd-viscous force biases the circulations in the two chambers differently:
    one vortex is amplified, while the other is completely suppressed. The background
    color represents local flow speed in a logarithmic scale, and the arrows indicate the local flow direction.
    }
    \label{fig:two-vortex-vs-one-vortex}
\end{figure}

In this work we show that a spatially varying Berry-curvature-induced odd viscosity provides a control knob for realizing the effects of odd viscosity on the flow profile. To demonstrate this effect in an experimentally accessible regime, we study the electronic flow in a two-chamber geometry~\cite{Zhang2026,AharonSteinberg2022}, summarized in Fig.~\ref{fig:two-vortex-vs-one-vortex}. In an ordinary viscous electron fluid, or in a fluid with spatially uniform Berry curvature, the two chambers host a pair of nearly mirror-symmetric counter-rotating vortices, illustrated in Fig.~\ref{fig:two-vortex-vs-one-vortex}(a). The flow pattern is set primarily by geometry: the injected current passes through the central constriction, while the side chambers support recirculating eddies. A spatial gradient of Berry curvature generates a gradient in the Hall viscosity, changing this situation qualitatively. The gradient acts as a chiral selector for the vortices, as the vortex in one chamber is enhanced, while the other vortex is weakened and can be completely suppressed (see Fig.~\ref{fig:two-vortex-vs-one-vortex}(b)). The result is a striking conversion of a symmetric two-vortex pattern into an asymmetric state with one amplified vortex and one extinguished vortex. We denote this as the `viscochiral' regime of transport.

This transition from two vortices to a single vortex is the smoking-gun signature we emphasize in this paper. It is not tied to the microscopic details of the particular device, nor to the precise chamber shape. The general mechanism is that a spatially varying Berry-curvature-induced odd viscosity removes vorticity from one chamber, while adding vorticity to the other. Therefore, whenever a hydrodynamic flow contains recirculating structures, a Hall viscosity gradient can amplify vortices of one chirality, and suppress vortices of the other.

The control parameter which determines the transition from the regular hydrodynamic regime to the viscochiral regime is set by the dimensionless parameter $\chi=\Delta \eta_H / \mu$, which is the ratio of the contrast in the Hall viscosity $\Delta \eta_H$, to the regular viscosity $\mu$.

The viscosity tensor relates the stress tensor with the gradients of the flow velocity,
\be \label{eq:pi-ij-definition}
\pi_{ij} = \eta_{ij kl} \partial_l u_k.
\ee

The breaking of time-reversal symmetry in metals can generate an odd component of the viscosity in the electronic flow, as demonstrated in Ref.~\cite{RaoBradlyn2020} using the Kubo formula~\cite{Kubo_statistical-mechanical_I_1957, Kubo_statistical-mechanical_II_1957}, and more recently in Ref.~\cite{SrivastavaMukerjee2025} using the semiclassical formalism constrained by the Onsager relations, based on the framework developed in Refs.~\cite{Panigrahi_energy_2023, Chadha_vortical_2025}. For an isotropic dispersion, this contribution to the viscosity tensor takes the form,
\be
\eta^{(\mathrm{odd})}_{ijkl}(\mathbf{r})=\delta_{ik}\veps_{jl}\,\etal(\mathbf{r}),
\label{eq:oddtensor}
\ee
where $\varepsilon_{ij}$ and $\delta_{ij}$ are the Levi-Civita tensor and the Kronecker delta tensor, respectively, and 
\be
\etal(\mathbf{r})=\hbar \sum_{Q=K,K'}\int_{\mathbf{k} \in Q}\!\frac{d^2 \vec k}{(2\pi)^2}\,\frac{k^2}{2}\,\Omega^z_Q({k})\,f_Q^0(\mathbf{r},k).
\label{eq:etaH_def}
\ee
Appendix A relates Eq.~\eqref{eq:oddtensor} to the more general odd-viscosity decomposition commonly used in the hydrodynamics literature.
Here $\Omega^z$ is the Berry curvature and $f^0$ the local equilibrium distribution. The index $Q$ accounts for the contribution from $K$ and $K'$ valleys. Clearly, $\eta_H = 0$ when time reversal symmetry is present, and the two valleys have equal population whereas it has opposite signs for valley-polarized phases. While this is a non-interacting formula, the main conclusions of the paper would not change for interacting systems because they are derived under the condition that electrons in $K$ and $K'$ valleys contribute to opposite signs of Hall viscosity. We focus on the case where the profile of Hall viscosity changes sharply at a domain boundary, as expected in realistic systems~\cite{Huang_private_communication},
\be
\etal(y)=\frac{\Delta \eta_H}{2} \tanh \left(\frac{y}{a}\right).
\label{eq:step_profile}
\ee

\begin{figure*}[t]
    \includegraphics[width=\linewidth]{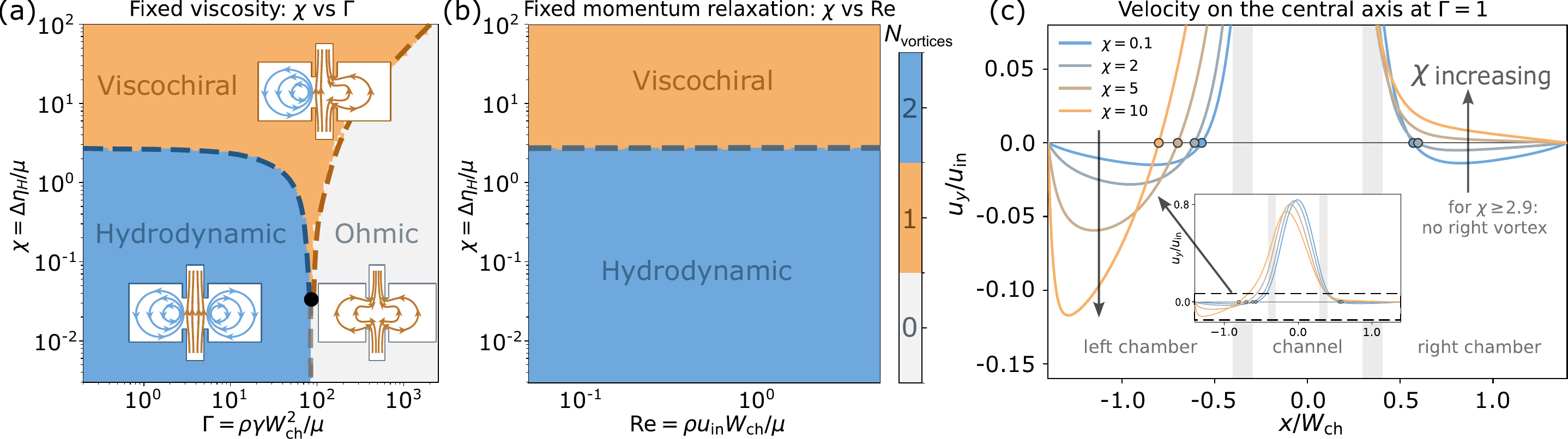}
    \caption{Phase diagram of the flow-profiles in two chambers: (a) Different regimes as a function of the dimensionless jump in Hall viscosity and the dimensionless Ohmic scattering rate $\Gamma = \rho \gamma W_{\rm ch}^2/\mu$. The phase diagram hosts three regimes, hydrodynamic (blue), with a symmetric pair of vortices of opposite signs, viscochiral (orange), in which one vortex is amplified and the other is extinguished, and Ohmic (gray), a potential flow lacking vortices. The schematics of the flow profile in each regime is shown in the insets. At small $\Gamma$ the transition between viscochiral regime and the hydrodynamic regime occurs at $\chi_c \approx 2.9$, whereas near the hydrodynamic-Ohmic boundary, the viscochiral phase extends down to $\chi \sim 0.05$ (the `viscochiral tongue'), because the vortices are already weakened by the Ohmic relaxation, and one of them is easily destroyed by the Hall viscosity. (b) The boundary between the viscochiral and hydrodynamic regimes as a function of $\chi$ and $\rm Re = \rho u_{\rm in} W_{\rm ch}/\mu$, at a small Ohmic damping $\Gamma$. Here the boundary is horizontal, implying that the regime of flow is determined by $\chi$ alone, and is independent of the flow velocity. The colorbar encodes the number of vortices in the two chambers. (c) The velocity profile along the direction of the main channel, on a line through both the chambers, at $\Gamma = 1$, for $\chi = 0.1,2,5,10$. As $\chi$ is increased, the vortex in the left chamber gets amplified, and the one on the right chamber gets destroyed, which is evident from the lack of sign change of velocity within the right chamber for $\chi \gtrsim 2.9$. The inset shows the velocity profile within the main channel.
    }
    \label{fig:chiral_selection}
\end{figure*}

Such a scenario can arise when the source and the drain of the device have opposite valley-polarization, and the domain boundary is near the chambers. The typical domain wall width is $a \sim 50 \rm{ nm}$~\cite{Huang_private_communication}. We note that the Hall viscosity profile in Eq.~\eqref{eq:step_profile} is just an example, while the physics of the viscochiral regime is agnostic to the details of the profile. For instance, we found that the viscochiral regime continues to exist when there is a uniform gradient of Hall viscosity, and the phase diagram in the parameter space (analog of Fig.~\ref{fig:chiral_selection}(a)) looks qualitatively similar.

\textit{Navier-Stokes equation modified by Hall viscosity gradient---} 
Coulomb interactions suppress long-wavelength density fluctuations in the charged electronic fluid, allowing us to treat the flow as effectively incompressible, $\nabla \cdot \mathbf{u}=0$.
The resulting steady Navier-Stokes equation is (see Appendix C),
\be
\rho(\mathbf{u}\cdot\nabla)\mathbf u
=
-\mathbf\nabla p
+\mu \nabla^2 \mathbf u
-\gamma \rho \mathbf u
+\mathbf F^{\mathrm{ext}}
-
\veps_{jl}\,\bigl(\p_j \etal\bigr)\,\p_l \mathbf u.
\label{eq:NS_main}
\ee
Here $\gamma$ is the Ohmic momentum relaxation rate.

Eq.~\eqref{eq:NS_main} immediately shows why a gradient of valley polarization rather than a uniform valley polarization is necessary. The new force term is controlled by the gradient $\partial_y \eta_H$ rather than the absolute value of $\eta_H$, and the gradient is zero when there is uniform valley polarization (or, when the system is unpolarized). The other possible spatial derivative term arising for a spatially uniform Hall viscosity is $(\varepsilon_{jl} \eta_H \partial_j \partial_l \vec u)$, which becomes identically zero due to the anti-symmetry of the Levi-Civita tensor.

To obtain the flow pattern in the chamber geometry, the Navier-Stokes equation (Eq.~\eqref{eq:NS_main}) is numerically solved within the chamber geometry in Fig.~\ref{fig:two-vortex-vs-one-vortex}, such that the flow velocity goes to zero at each boundary, and a current is injected into the channel such that the flow has a parabolic profile near the entrance of the channel. 
To quantitatively identify vortices, we compute the dimensionless quantity $\frac{|\oint \vec{u}\cdot \mathrm{d}\vec l|}{\oint |\vec{u}| \mathrm{d} l}$ evaluated along a closed loop parallel to the boundary of the chamber. When this quantity is larger than a threshold, it is taken as a signature of a vortex.
We also enforce the condition that the flow velocity for a vortex must be strong enough, so as to eliminate fictitious vortices emerging from numerical noise. In the viscochiral regime, the sign of $\Delta \eta_H$ selects which chamber of the device will host the vortex.

The effect of the Hall viscosity on the vortex in the chamber $\mathcal D$ (either the left or right chamber) can be understood by studying the dynamic effect of suddenly turning on the Hall viscosity gradient term on a steady state system without the Hall viscosity gradient. We consider the vorticity equation describing the dynamics of vorticity $\omega= (\nabla \times \vec u)_z$, yielding,
\begin{equation}\label{eq:vorticity-eqn}
    {\partial_t \omega} + (\vec u \cdot \nabla) \omega = \frac{\mu}{\rho} \nabla^2 \omega - \gamma \omega + \frac{\tau}{\rho}  + \frac{1}{\rho} (\partial_y \eta_H) \partial_x \omega,
\end{equation}
with $\tau = (\nabla \times \vec F^{\rm ext})_z$. Here we neglect the term $(\partial_y^2 \eta_H) \partial_x u_x$, because in the region where odd viscosity has a strong position dependence, the velocity along $x$ is negligible, and away from this region, $\partial_y^2 \eta_H$ is zero.
Since the system was in a steady state, initially the time derivative will be balanced by the Hall viscosity term, while all other terms will cancel each other. Multiplying both sides by $\omega$ and integrating over the whole area of chamber $\mathcal D$, we obtain,
\begin{equation}\label{eq:initial-time-evolution}
\begin{aligned}
    \frac{d}{dt} \left(\int_{\mathcal D} {\omega^2} d^2 \vec r \right) &= \int_{\mathcal D} \frac{1}{\rho} (\partial_y \eta_H) \partial_x (\omega^2) d^2 \vec r \\
    &\approx \frac{\Delta \eta_H}{\rho} \left[\left . \omega^2 \right |_{\rm right} - \left . \omega^2 \right|_{\rm left}\right],
\end{aligned}
\end{equation}
where the subscripts `left' and `right' correspond to the values at the left and right boundaries of the chamber.

Since the vorticity is mostly confined near the neck of the chambers, and goes to zero near the edges, the right hand side of Eq.~\eqref{eq:initial-time-evolution} is positive for the chamber on the left, and is negative for the chamber on the right. Consequently, the mean-square vorticity within a chamber (quantitatively measured by enstrophy $\int_{\mathcal D} {\omega^2} d^2 \vec r$ within the chamber) rises in the chamber on the left and decreases in the chamber on the right. As a result, the magnitude of vorticity will be amplified in the chamber on the left, and it will diminish in the chamber on the right, and eventually, the system will reach an equilibrium with an asymmetric distribution of vortices. It is interesting to note that the Hall viscosity does not directly couple to the sign of the vorticity, but rather, it transports the magnitude of vorticity in and out of the left and right chambers, respectively. Moreover, as long as the width of the domain wall is less than that of the vortex core, which is expected in realistic systems, the result in Eq.~\eqref{eq:initial-time-evolution} is independent of the domain wall width. In case the domain wall is very wide, only the variation of odd viscosity across the lengthscale of the vortex core will significantly contribute. We also perturbatively solved the effect of the odd-viscosity-gradient term on localized vortices, and found that it spatially shifts vortices of either orientation in the same direction (see Appendix E). Consequently, the odd-viscosity-gradient term pushes one of the vortices from the chamber into the main channel, effectively destroying it. Even when the Hall viscosity does not sharply change, the sign of the time-evolution in enstrophy will remain the same in Eq.~\eqref{eq:initial-time-evolution}, still causing a similar asymmetry in vortex strength.

\begin{figure}
    \includegraphics[width=0.9\linewidth]{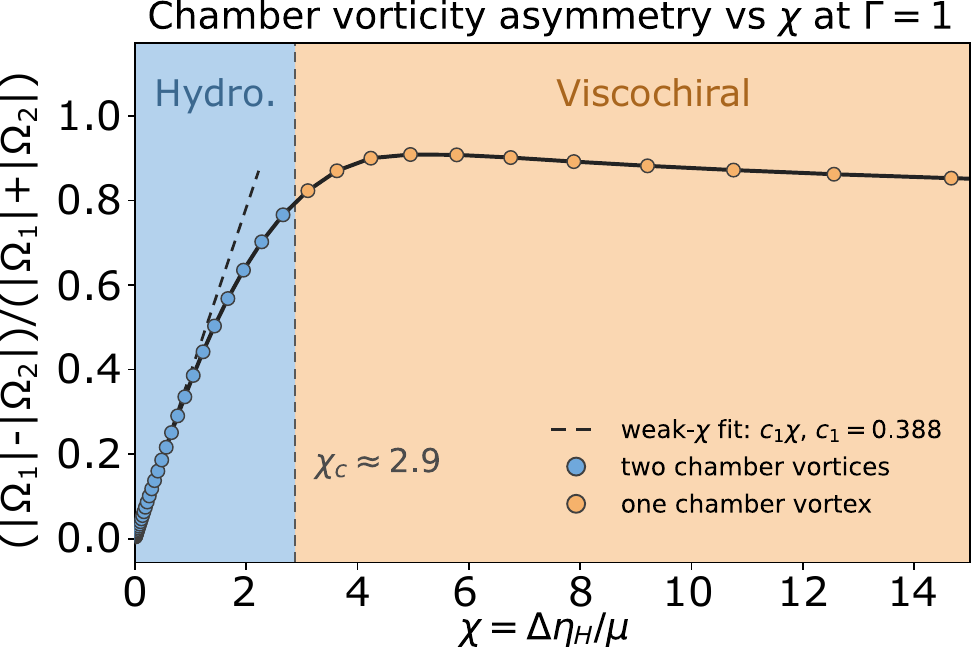}
    \caption{ The relative asymmetry in the strength of vorticity in the two chambers as a function of the dimensionless jump in Hall viscosity, $\chi$, for a fixed value of Ohmic scattering rate. Here, the magnitude of the vorticity in a chamber is defined as $\Omega_{1,2} = \int_{\mathcal D_{1,2}} \omega(\vec r) d^2 \vec r$.
    The transition to the viscochiral regime is numerically found to occur when $\chi_c \approx 2.9$ beyond which one vortex gets completely suppressed. In comparison, in the `viscochiral tongue' region of the phase diagram presented in Fig.~\ref{fig:chiral_selection}(a), the transition occurs at a much lower critical value of $\chi_c \approx 0.05$. We note that the linear fit at small values of $\chi$ estimates the numerical value of $\chi_c$ quite accurately.
    }
    \label{fig:vorticity_asymmetry}
\end{figure}

\textit{Parameter space phase diagram for three regimes of transport---} The phase diagram of the flow profile for different regimes of transport, as a function of the jump in dimensionless Hall viscosity $\chi (=\Delta \eta_H/\mu)$ and dimensionless Ohmic scattering rate $\Gamma$, was computed by numerically solving the Navier-Stokes equation in the two-chamber geometry, and it is presented in Fig.~\ref{fig:chiral_selection}(a).
There are three regimes, namely, hydrodynamic (coded with blue), Ohmic (coded with gray), and viscochiral (coded with orange).
When both $\Gamma$ and $\chi$ are small, the system is in the hydrodynamic regime, featuring two vortices in the two chambers. As the Ohmic dissipation rate $\Gamma$ is increased (while keeping the Hall viscosity gradient fixed), the system eventually transitions to the Ohmic regime where the velocity is described by a potential flow, and the system does not feature any vortices (see Appendix D for an illustration of typical flow profiles). These features were experimentally observed in Ref.~\cite{Zhang2026}.
At small values of Ohmic relaxation rate, when the jump in Hall viscosity is increased, the system enters the viscochiral regime, where one vortex is heavily amplified compared to the other one. The existence of this regime is the central result of this paper. The dependence of the relative strength of the vorticity in the two chambers as a function of $\chi$ is presented in Fig.~\ref{fig:vorticity_asymmetry}, and the local flow profile on a line passing through both the chambers is presented in Fig.~\ref{fig:chiral_selection}(c).

A very interesting situation occurs when the momentum-relaxation rate is large, near the verge of entering the Ohmic regime. In this case, the viscochiral regime can be stabilized at a much smaller value of Hall viscosity gradient, and this value ($\chi_c \approx 0.05$) is nearly two orders of magnitude smaller than the critical value of Hall viscosity gradient at small momentum relaxation rate ($\chi_c \approx 2.9$), creating a `viscochiral tongue' in the phase diagram. This can be qualitatively understood as follows. When the momentum relaxation rate is large such that the system is about to enter the Ohmic regime, the vortices are weakened. In this scenario, a relatively small value of Hall viscosity gradient can significantly alter the relative strength of the two vortices, as one of them will be weakened further and be completely destroyed, while the other will be amplified. As a result, in this regime, a small Hall viscosity gradient will significantly enhance the ratio of the vorticity in the two chambers, resulting in the `viscochiral tongue' region. We find that the `viscochiral tongue' exists even for a uniform (rather than localized) gradient of Hall viscosity, and the phase diagram remains qualitatively similar.

The behavior of the system as a function of the dimensionless Hall viscosity gradient and the dimensionless flow velocity is also studied  (Fig.~\ref{fig:chiral_selection}(b)) in the regime of a very low momentum relaxation rate $\gamma$. The flow velocity is non-dimensionalized by considering the Reynolds number ${\rm{Re}} = \rho u_{\rm in} W_{\rm ch}/\mu$, where $W_{\rm ch}$ is the width of the channel. It is found that for a fixed viscosity $\mu$, varying the Reynolds number by changing the flow velocity $u_{\rm in}$ does not alter the regime of flow, as evident from the horizontal boundary between the two regimes. The viscochiral or hydrodynamic nature of the flow only depends on the dimensionless quantity $\chi={\Delta \eta_H}/{\mu}$, which encodes the strength of the jump in Hall viscosity compared to the regular viscosity.

\textit{Practical considerations---} Now we discuss how the viscochiral regime may be experimentally realized in systems with broken time reversal symmetry. Vortices in a double chamber geometry were realized with bilayer graphene in Ref.~\cite{Zhang2026}, and the transition between Ohmic and hydrodynamic regimes was observed. The hydrodynamic regime requires electron-electron mean-free path to be substantially smaller than the system size, and in Ref.~\cite{Zhang2026}, it was estimated to be $30$--$50 \rm nm$. At low temperatures and carrier densities, and under high displacement fields, Bernal bilayer graphene hosts valley-polarized phases with spontaneously broken time-reversal symmetry. It has been experimentally demonstrated that under a high displacement field, and at low electronic density, the valley polarization can be locally controlled by an out-of-plane magnetic field~\cite{Zhou_half-_2021, Han_orbital_2023, Winterer_ferroelectric_2024}, as the orbital magnetization interacts with the magnetic field via a Zeeman-like coupling, and the two valleys have opposite signs of orbital magnetization. Applying opposite out-of-plane magnetic fields at the source and the drain will stabilize different valley polarizations at the two sides of the system, generating a Hall viscosity gradient, which will favor the viscochiral regime of transport discussed earlier. Of course, the viscochiral effect will not be observed if the sharp domain boundary falls away from the chambers. It may require some engineering with an external magnetic field to make the domain boundary align with the entrance of the chambers.
For an electronic density of $n=3 \times 10^{11}/\rm{cm}^2$, with an effective screened potential difference $D=60 {\text{ meV}}$ between the layers, we find $\chi = \Delta \eta_H/\mu \sim 0.33$ (see Appendix B for details), which corresponds to the viscochiral regime in the `viscochiral tongue' region in the phase diagram shown in Fig.~\ref{fig:chiral_selection}(a).
Rhombohedral multilayers host a greater magnitude of Berry phase per valley, and may provide an enhanced value of $\chi$.
Moreover, since the shear viscosity of Fermi liquids has a temperature dependence $\mu \sim 1/T^2$, the parameter $\chi$ can be amplified by increasing the temperature as long as the temperature remains below the critical temperature for hosting the valley-polarized phase. A similar effect is expected to show up if the source (or the drain) is valley-polarized above the critical temperature with an external magnetic field, but the drain (or the source) is still valley-unpolarized.

\textit{Discussion---}
We studied the effects of a Hall viscosity gradient generated by a spatially varying Berry curvature on the hydrodynamic flow of electrons in a channel with two chamber geometry. Without any Hall viscosity gradient, the system hosts two vortices of equal magnitude within the two chambers in the limit of small Ohmic dissipation rate, which are destroyed to give rise to a potential flow when the Ohmic dissipation is increased. While a uniform Hall viscosity does not alter the flow described by the Navier-Stokes equation, a strong gradient in Hall viscosity causes an asymmetry in the strength of the two vortices, giving rise to the viscochiral regime. The dimensionless parameter describing the transition between the viscochiral regime and the hydrodynamic regime is $\chi=\Delta \eta_H/\mu$, the ratio of the jump in Hall viscosity to the shear viscosity of the system. At larger Ohmic dissipation rates, the stronger vortex in the viscochiral regime gets destroyed, giving way to the vortex-free Ohmic regime. An interesting phenomenon occurs when the Ohmic dissipation is increased near the critical value where the viscochiral regime eventually turns into Ohmic flow. Here, the vortex already weakened by the Ohmic dissipation gets further weakened by the Hall viscosity gradient, generating a viscochiral flow at a much smaller value of the dimensionless parameter $\chi$, and featuring a `viscochiral tongue' region in the phase diagram (Fig.~\ref{fig:chiral_selection}(a)). The existence of the viscochiral regime and the `viscochiral tongue' in the phase diagram are the main results of this paper.
The two-vortex structure in a double chamber geometry is known to be robust consequence of the recirculating flow, independent of the shape of the device~\cite{AharonSteinberg2022, Zhang2026, NazaryanLevitov2024}, and the vorticity imbalance follows from the sign structure of the right hand side in Eq.~\eqref{eq:initial-time-evolution} rather than the specific profile of Hall viscosity assumed in Eq.~\eqref{eq:step_profile}. Therefore, we expect the viscochiral regime to be a generic feature of hydrodynamic transport under strong enough Hall viscosity gradient.

Since Hall viscosity is odd under time-reversal~\cite{RaoBradlyn2020, SrivastavaMukerjee2025}, and $K$ and $K'$ valleys of rhombohedral graphene host opposite signs of Berry curvature, a Hall viscosity gradient will be generated at the boundary between domains of two valley-polarized phases. Based on the recent experiments in Ref.~\cite{Zhang2026}, we discuss a realistic geometry where such a domain wall can be experimentally realized. At a temperature low enough to host valley-polarized phases the source and drain can be driven into opposite valley polarized states by applying a local out-of-plane magnetic field of opposite signs, which will generate a domain boundary with concentrated Hall viscosity gradient, potentially realizing the viscochiral regime of transport. We argue that the `viscochiral tongue' regime of the phase diagram can be experimentally realized with experimentally realizable system parameters in Bernal bilayer graphene, and may be directly realized in experiments similar to that in Ref.~\cite{Zhang2026}. The current flow profiles can be reconstructed from the out-of-plane magnetic field profile measured with a superconducting quantum interference device (SQUID)~\cite{Vasyukov_a_2013, Finkler_self-aligned_2010}, as was done in Ref.~\cite{Zhang2026}.
Outside graphene-based systems, a non-uniform Hall viscosity may also be engineered with an appropriate non-uniform strain in flexomagnetic systems~\cite{Lukashev_flexomagnetic_2010, Tang_flexomagnetism_2025}. Since the double vortex structure is known to generic and independent of the exact details of the chamber geometry~\cite{AharonSteinberg2022, Zhang2026, NazaryanLevitov2024}, and the asymmetry of vortices Eq.~\eqref{eq:initial-time-evolution} is independent of the details of the spatial profile of the Hall viscosity gradient, the `viscochiral regime' is expected to be a generic transport feature with a strong enough Hall viscosity gradient.

\textit{Note added---}
During the completion of this manuscript, we became aware of Ref.~\cite{Huang_private_communication}, which carried out a microscopic calculation and found that the domain-wall width in rhombohedral graphene is only a few interparticle spacings, placing it in the regime relevant to our theory. We also became aware of Refs.~\cite{Huang_private_communication2, Huang_private_communication3}, which reported interesting transport signatures associated with such domain walls.

\textit{Code availability---}
The code used in this work will be made publicly available prior to the publication.

\textit{Acknowledgments---} We thank Nisarg Chadha, Rafael Fernandes, Chunli Huang, Daniel Kaplan, Subroto Mukerjee, Yugo Onishi, Vladislav Poliakov and Anubhav Srivastava for insightful discussions. We acknowledge Leonid Levitov for bringing our attention to the experiment in Ref.~\cite{Zhang2026}.

\bibliography{references_v1.bib}

\onecolumngrid
\appendix
\numberwithin{equation}{section}

\section{Relation between general form of viscosity tensor and the specific form obtained in Ref.~\cite{SrivastavaMukerjee2025}}

The most general isotropic 2D viscosity tensor can be written as~\cite{Fruchart2023},
\begin{equation}
\eta_{ijkl}
= \zeta \, \delta_{ij}\delta_{kl}
- \eta_A \, \varepsilon_{ij}\delta_{kl}
- \eta_B \, \delta_{ij}\varepsilon_{kl}
+ \eta_R \, \varepsilon_{ij}\varepsilon_{kl}
+ \mu\,(\delta_{ik}\delta_{jl}+\delta_{il}\delta_{jk}-\delta_{ij}\delta_{kl})
+ \eta_o\,({\varepsilon_{ik}\delta_{jl}}+{ \varepsilon_{jl}\delta_{ik}}).
\label{eq:annualreview}
\end{equation}

Ref.~\cite{SrivastavaMukerjee2025} obtains an odd viscosity tensor of the form,
\begin{equation}
\eta^{(\mathrm{odd})}_{ijkl}(\vec r) = -L^{9,\rm{(odd)}}_{ij,kl} = \eta_H (\vec r){\delta_{ik}\varepsilon_{jl}},
\label{eq:berrytensor}
\end{equation}
with
\be
\eta_H(\mathbf{r})=\hbar \sum_{Q=K,K'}\int_{\mathbf{k} \in Q}\!\frac{d^2k}{(2\pi)^2}\,\frac{k^2}{2}\,\Omega^z_Q({k})\,f_Q^0(\mathbf{r},k)
\label{eq:C-def}
\ee

Using the following tensor relation valid in two dimensions, 
\be
\delta_{ik}\varepsilon_{jl}
=
\frac12\bigl(\varepsilon_{ik}\delta_{jl}+\varepsilon_{jl}\delta_{ik}\bigr)
-\frac12\varepsilon_{ij}\delta_{kl}
+\frac12\delta_{ij}\varepsilon_{kl},\ee
we obtain,
\begin{align}\label{eq:app-odd-viscosity-details}
\eta^{(\mathrm{odd})}_{ijkl}
&= \eta_H \left[
\frac12\bigl(\varepsilon_{ik}\delta_{jl}+\varepsilon_{jl}\delta_{ik}\bigr)
-\frac12\varepsilon_{ij}\delta_{kl}
+\frac12\delta_{ij}\varepsilon_{kl}
\right].
\end{align}
Comparing Eq.~\eqref{eq:app-odd-viscosity-details} term by term with Eq.~\eqref{eq:annualreview}, we obtain,
\begin{equation}
\eta_o = \frac{\eta_H}{2}, \qquad
\eta_A = \frac{\eta_H}{2}, \qquad
\eta_B = -\frac{\eta_H}{2}.
\end{equation}

This odd viscosity term does not contribute to the remaining viscous coefficients $\mu$ and $\zeta$. In Appendix C, we show that a spatially uniform odd viscosity has no effect on the flow profile.

\section{Estimation of Hall viscosity in bilayer graphene}
We consider the effective two-band Hamiltonian for Bernal bilayer graphene,
\begin{equation}
    H(\mathbf{p}) = \begin{pmatrix}
\frac{D}{2} & \frac{(p_x-i p_y)^2}{2 m^*} \\
\frac{(p_x+i p_y)^2}{2 m^*} & -\frac{D}{2} 
\end{pmatrix}
\end{equation}
with $m^* = \frac{\gamma_1}{2 v^2} = 0.034 m_e$.

The Berry curvature is given by
\begin{equation}
    \Omega_z(k) = \frac{\frac{D}{2} \frac{p^2}{{m^*}^2}}{2 \left[(\frac{D}{2})^2 + (\frac{p^2}{2 m^*})^2\right]^{\frac{3}{2}}}
\end{equation}

The Hall viscosity of a fully spin-valley polarized phase is 
\begin{equation}
\begin{aligned}
    \etal&=\hbar \int_{k<k_F}\!\frac{d^2k}{(2\pi)^2}\,\frac{k^2}{2}\,\Omega_z(|\mathbf{k}|)\\
    &= n\hbar \frac{D m^*}{4\pi n \hbar^2} \left[\sinh^{-1}\left(\frac{4 \pi n \hbar^2}{m^* D}\right) - \frac{\frac{4\pi n \hbar^2}{m^* D}}{\sqrt{1 + (\frac{4\pi n \hbar^2}{m^* D})^2}} \right].
\end{aligned}
\end{equation}

This expression is maximum when $\frac{D m^*}{4\pi n \hbar^2} = 0.3424$, with the maximum value being $0.29 n \hbar$. A typical density is $n = 3\times 10^{11} \rm{cm}^{-2}$ and a typical displacement field $D = 60 \rm{meV}$. For these, the dimensionless parameter takes the value $\frac{D m^*}{4\pi n \hbar^2}=0.71$ and $\eta_H = 0.23 n \hbar$.

For density $n = 3\times 10^{11} \rm{cm}^{-2}$, the Fermi wavenumber is $k_F = 0.19/\rm{nm}$.

The chamber-imaging experiment of Ref.~\cite{Zhang2026} finds the representative best-fit value $\ell_{ee}=29\,{\rm nm}$.  Then, $\mu \approx \frac{1}{4} n\hbar k_F l_{ee} = 1.37 n\hbar$.

Then, $\eta_H/\mu = 0.167$ for the valley polarized phase. Across a domain wall between two valley-polarized phases, the value of the difference in Hall viscosity $\Delta \eta_H$ will be twice as large, and $\chi=\Delta \eta_H/\mu = 0.33$.

Since the viscosity of a Fermi liquid decreases with rising temperature as $\mu \sim 1/T^2$, the ratio $\Delta \eta_H/\mu$ can be even larger at a slightly higher temperature, as long as the system is in the valley-polarized regime.

\section{Derivation of the Navier-Stokes equation with Hall viscosity term and absence of a bulk force for uniform $\eta_H$}
The component-wise form of the Navier-Stokes equation with a phenomenological momentum relaxation term and an odd term in the stress tensor is,
\be
\rho \partial_t u_i + \rho({\vec u}\!\cdot\!\nabla)u_i
=
-\mathbf\partial_i p
+\mu \nabla^2 u_i
-\gamma \rho u_i
+F_i^{\mathrm{ext}}
-\partial_j \pi^{\rm (odd)}_{ij}.
\label{eq:full-NS}
\ee

Using the relations $\pi_{ij} = \eta_{ijkl}\partial_l u_k$ and $\eta^{\rm (odd)}_{ijkl} (\mathbf{r})=\delta_{ik}\veps_{jl}\,\etal(\mathbf{r})$, we obtain Eq.~\eqref{eq:NS_main} in the main text. It is to be noted that when $\etal(\mathbf{r})$ is spatially uniform, $\partial_j \pi^{\rm (odd)}_{ij} =  \eta_H\varepsilon_{jl} \partial_j \partial_l u_i = 0$, since the Levi-Civita tensor $\varepsilon_{jl}$ is anti-symmetric while the product of two derivatives $\partial_j \partial_l$ is symmetric.

\section{Vortex-free flow profiles in the Ohmic regime}
\begin{figure}[h]
    \centering
    \includegraphics[width=0.3\linewidth]{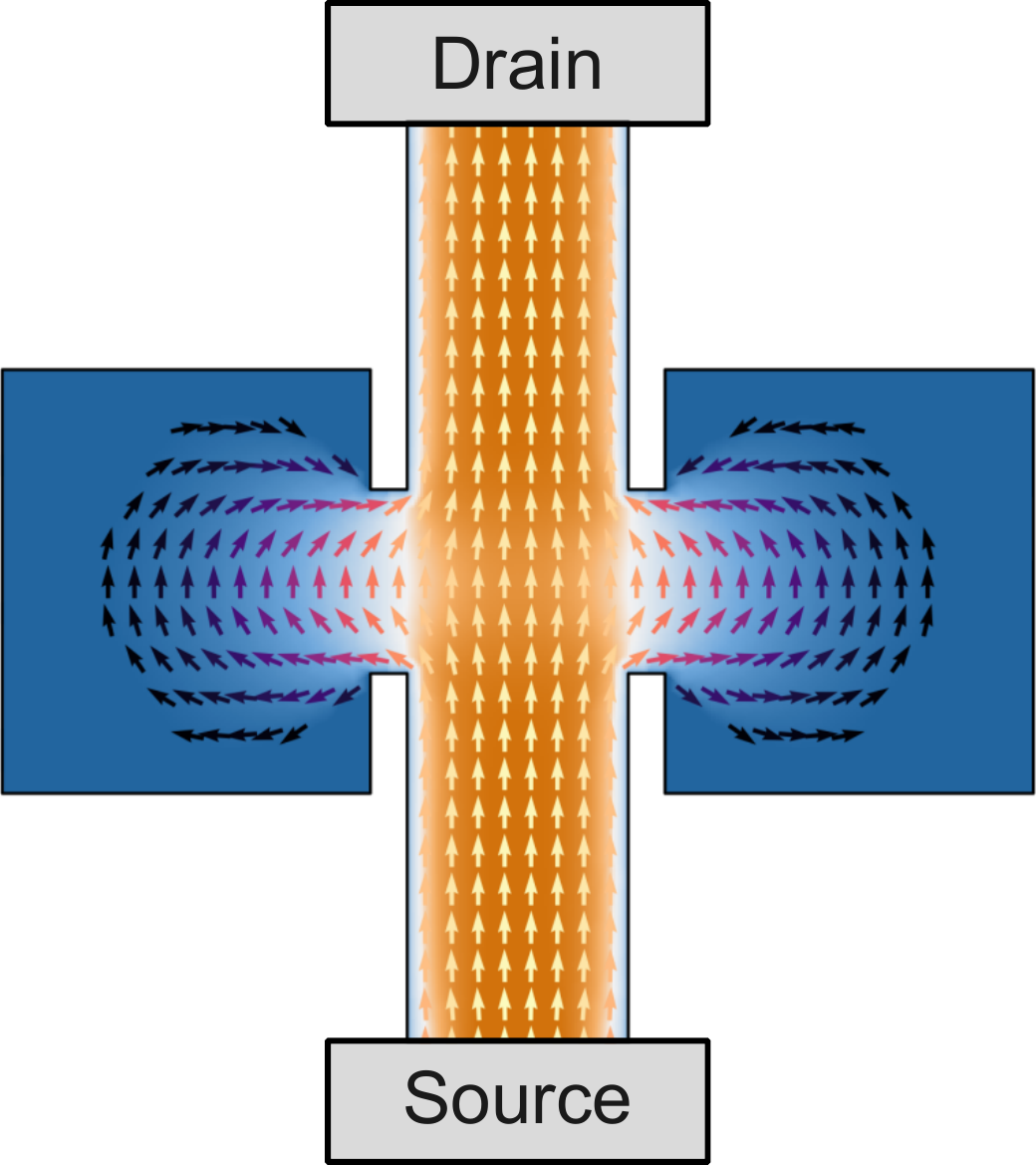}
    \caption{A typical flow profile in the Ohmic regime, lacking vortices in either chamber.}
    \label{fig:Ohmic-profile}
\end{figure}

In the Ohmic regime, the current follows the electric field, which is given by the gradient of the potential.

Since $\vec j = \sigma \vec E$, and $\nabla \times \vec E = 0$ in the steady state, we must have $\nabla \times \vec j = 0$, which implies that the flow profile $\vec u$ cannot have any vortices, as illustrated in Fig.~\ref{fig:Ohmic-profile}.

\section{Effects of Hall viscosity gradient on a localized vortex}

The steady-state vorticity equation with a uniform gradient of Hall viscosity, $\frac{1}{\rho}\partial_y \eta_H = C$ is,
\begin{equation}
    (\vec u \cdot \nabla) \omega = \frac{\mu}{\rho}\nabla^2 \omega + (\nabla \times \vec F)_z + C \partial_x \omega.
\end{equation}

We consider a localized Gaussian vorticity profile,
\begin{equation}
    \omega_0 (\vec r) = \frac{\Omega}{2 \pi a^2} e^{-r^2/(2a^2)},
\end{equation}
which has total vorticity $\int \omega(\vec r) d^2 \vec r = \Omega$. We want to study the effect of a non-zero gradient of Hall viscosity on it. Substituting this vorticity profile in the vorticity equation, it can be shown that this vortex is stabilized for $C=0$ by a force field
\begin{equation}
    \vec F = \frac{\mu}{\rho} \frac{\Omega r}{2 \pi a^4} e^{-r^2/(2 a^2)} \hat{\theta}.
\end{equation}

To study the effect of the Hall-viscosity gradient term, we solve the vorticity equation perturbatively in the limit of small $C$, and write $\omega(\vec r) = \omega_0 (\vec r) + C \omega_1(\vec r) + \mathcal{O}(C^2)$. Substituting this ansatz in the vorticity equation, we obtain,
\begin{equation}
    \frac{\mu}{\rho} \nabla^2(C \omega_1) \approx - C \partial_x \omega_0.
\end{equation}

Substituting the form of $\omega_0$ in this equation, we find the solution, $\omega_1 = -\frac{\rho}{\mu} \frac{\Omega}{2\pi} \frac{x}{r^2} \left(1 - e^{-r^2/(2 a^2)} \right)$. The full expression for vorticity is,
\begin{equation}
\begin{aligned}
    \omega(\vec r) &= \frac{\Omega}{2 \pi a^2}  \left[e^{-r^2/(2 a^2)} - C \frac{\rho}{\mu} \frac{x a^2}{r^2} \left(1- e^{-r^2/(2 a^2)}\right) \right]\\
    & \approx \frac{\Omega}{2 \pi a^2} \left[1 -\frac{C \rho x}{2\mu} - \frac{x^2 + y^2}{2a^2}\right]
\end{aligned}
\end{equation}
where in the last line we expanded the Gaussian functions in their Taylor series for small values of $r$.

The vorticity is maximized at $(x^* = -\frac{C \rho a^2}{2\mu}, y^*=0)$, which means, the vortex gets shifted towards left for a positive $C$, and this result does not depend on the sign of its vorticity $\Omega$.

This explains why the vortex on the chamber in the right gets pulled into the main channel, and gets destroyed, while the vortex in the left chamber survives. The opposite effect will occur when the sign of the gradient of Hall viscosity is flipped.

\end{document}